\documentclass[pra,aps,twocolumn,10pt,showpacs,groupedaddress,superscriptaddress,floatfix]{revtex4-1}
\usepackage{amsmath,amsfonts,amssymb,graphics,graphicx,epsfig,color,times}%,bbm}
\usepackage[utf8x]{inputenc}
\usepackage{slashbox}
\usepackage{color}
\usepackage{bbm, dsfont} 
\usepackage{subfigure}
\usepackage{hyperref}
\usepackage{mathrsfs}
\usepackage{verbatim}
\begin{document}
\title{Spin-incoherent one-dimensional spin-1 Bose Luttinger liquid}
\author{H. H. Jen}
\affiliation{Institute of Physics, Academia Sinica, Taipei 11529, Taiwan}
\author{S.-K. Yip}
%\noaffiliation
\affiliation{Institute of Physics, Academia Sinica, Taipei 11529, Taiwan}
\affiliation{Institute of Atomic and Molecular Sciences, Academia Sinica, Taipei 10617, Taiwan}

\date{\today}

\renewcommand{\r}{\mathbf{r}}
\newcommand{\f}{\mathbf{f}}

\def\be{\begin{align}}
\def\ee{\end{align}}
\def\bea{\begin{eqnarray}}
\def\eea{\end{eqnarray}}
\def\ba{\begin{array}}
\def\ea{\end{array}}
\def\bdm{\begin{displaymath}}
\def\edm{\end{displaymath}}
\def\red{\color{red}}

\begin{abstract}
We investigate spin-incoherent Luttinger liquid of a one-dimensional spin-1 Bose gas in a harmonic trap.\ In this regime highly degenerate spin configurations emerge since the spin exchange energy is much less than the thermal energy of the system, while the temperature is low enough that the lowest energetic orbitals are occupied.\ As an example we numerically study the momentum distribution of a one-dimensional spin-1 Bose gas in Tonks-Girardeau gas limit and in the sector of zero magnetization.\ We find that the momentum distributions broaden as the number of atoms increase due to the averaging of spin function overlaps.\ Large momentum ($p$) asymptotic is analytically derived, showing the universal $1/p^4$ dependence.\ We demonstrate that the spin-incoherent Luttinger liquid has a momentum distribution also distinct from spinless bosons at finite temperature.
\end{abstract}
\maketitle
\section{Introduction}
The interacting one-dimensional (1D) quantum system \cite{Giamarchi2004} can be described by a low-energy effective theory, so-called Luttinger liquid theory \cite{Haldane1981}.\ For spinful systems, one has spin-charge separation but both possess power-law decay of correlation functions.\ An interesting regime is spin-incoherent Luttinger liquid (SILL) \cite{Fiete2007} when the temperature is larger than the spin excitation energy while smaller than the charge excitation one.\ The exponential factor in the single-particle Green's function of SILL puts itself in a different universal class from the Luttinger liquid, and rich physics appears due to the highly degenerate spin Hamiltonian.\ The semiconductor quantum wire can be a platform to reach SILL regime.\ In this regime the spin Hamiltonian becomes irrelevant and negligible in the correlation functions that the spin degrees of freedom are non-propagating \cite{Fiete2007}.\ The spin-charge separation still holds simply because the charge velocity is much larger than the spin velocity.\ In this paper we consider the interacting 1D Bose gas \cite{Cazalilla2011} in the spin-incoherent regime.\ The dispersion relation of the density mode in a spin-1 Bose gas is linear \cite{Ho1998}, which is the same as the collective charge excitations in the electronic spin-$1/2$ system.\ As we will explain later, the spin velocity is much less than the sound velocity, therefore we also have the spin-incoherent regime if the temperature is higher than the spin excitation energy.\ We expect that the spin-incoherent 1D Bose gas is even better to demonstrate SILL physics due to the controllability of spatial dimensions and atom-atom interactions.

The strongly interacting and spinless bosonic particles have been experimentally realized in 1D tightly confined systems \cite{Paredes2004, Kinoshita2004, Haller2009}.\ These hard-core bosons become fermionized in the Tonks-Girardeau (TG) \cite{Tonks1936, Girardeau1960} gas limit that corresponds to infinitely strong atom-atom interaction, and is one of the few examples that have the exact solutions.\ In this limit their wavefunctions can be written down from the constraints that bosons are impenetrable and of bosonic symmetry.\ In the past two decades many investigations along this line involve the ground state properties of 1D hard-core bosons \cite{Girardeau2001} and strongly interacting Bose-Fermi mixtures \cite{Girardeau2007} in a harmonic trap, the momentum distribution \cite{Minguzzi2002, Olshanii2003, Xu2015}, quantum magnetism in 1D spinor Bose gas \cite{Deuretzbacher2008, Deuretzbacher2014, Volosniev2014, Yang2015, Deuretzbacher2016}, and using group theoretical methods in studying the ground states of spin-$1/2$ fermions \cite{Harshman2014} and the permutation symmetry of spinor quantum gases \cite{Yurovsky2014}.\ The experimental measurements of such quantum systems rely on the matter wave interferences, for example, the time-of-flight experiment using the focusing technique \cite{Shvarchuck2002, Davis2012, Jacqmin2012, Fang2016}, and the Bragg scattering spectroscopy \cite{Kozuma1999, Stenger1999, Steinhauer2002, Papp2008, Veeravalli2008, Pino2011}.\ In this paper we investigate the ground state properties of a 1D spin-1 Bose gas in SILL regime that has not been studied before, and we numerically calculate the momentum distributions that can be observed in experiments.\ We note that the momentum distributions of spin-$1/2$ bosons and fermions in SILL regime have been investigated in a homogeneous system \cite{Cheianov2005}.

The rest of the paper is organized as follows.\ In Sec. II we introduce the Hamiltonian of 1D spin-1 Bose gas and the general wavefunction with spin degree of freedom.\ In Sec. III, we consider 1D spin-1 Bose gas in TG limit.\ We derive the general form of density matrix with spin function overlaps assisted by the conjugacy classes of the permutation groups.\ An analytical derivation of large momentum asymptotic is demonstrated in Sec. IV, and we show the numerically calculated momentum distributions in Sec. V.\ We conclude in Sec. VI, and the Appendix has the technical details of the traces of various group elements which are used in determining the spin function overlaps.

%%%%%%%%%%%%%%%%%%%%%%%%%%%%%%%%%%%%%%%%%%%%%%%%%%%%%%%%%%%%%%%%%%%%%%%%%%%%%%%%
\section{Hamiltonian and wavefunction}
Starting from the Hamiltonian of ultracold spin-1 Bose gas in one dimension at zero magnetic field \cite{Ho1998, Ohmi1998, Deuretzbacher2008},
\bea
H&=&\sum_{i=1}^N\left[-\frac{\hbar^2}{2m}\frac{\partial^2}{\partial x_i^2}+\frac{1}{2}m\omega^2x_i^2\right]\mathbb{I}_{\rm spin}\nonumber\\
&+&\sum_{i<j}\delta(x_i-x_j)\left[U_0\mathbb{I}_{\rm spin}+U_2\f_i\cdot\f_j\right],
\eea
where $\f_i$ is the spin operator, $\mathbb{I}_{\rm spin}$ is the identity operator in spin space, $m$ is the mass of the bosons, $\omega$ is the axial trap frequency, and $U_{0,2}$ are the coupling constants of spin-independent and spin-dependent interactions respectively.\ For a highly elongated trap, the transverse trap frequency ($\omega_\perp$) is much larger than $\omega$ that the coupling constants effectively are $U_i$ $=$ $2\hbar\omega_\perp c_i$ with $c_0$ $=$ $(\tilde{a}_0+2\tilde{a}_2)/3$, and $c_2$ $=$ $(\tilde{a}_2-\tilde{a}_0)/3$, where $\tilde{a}_{0,2}$ $=$ $a_{0,2}/[1-1.46a_{0,2}/(\sqrt{2}l_\perp)]$ \cite{Olshanii1998, Yurovsky2008} with s-wave scattering lengths $a_{0,2}$, and $l_\perp$ $=$ $\sqrt{\hbar/(m\omega_\perp)}$ \cite{Ho1998}.\ 

In TG gas limit of scalar bosons, the system becomes fermionized that bosons do not penetrate each other, and their wavefunctions take the form of noninteracting fermions.\ For a spin-1 Bose gas with an infinite atom-atom interaction $U_0$ in a harmonic trap, we consider a SILL regime \cite{Fiete2007} where the spin exchange energy is much less than $k_BT$ where $k_B$ is the Boltzmann constant and $T$ is the temperature of the system, while the temperature is low enough that the system still occupies the orbitals in the lowest energy.\ The regime in general can be reached since $a_2$ $\approx$ $a_0$ for $^{23}$Na, $^{41}$K, and $^{87}$Rb that the interaction becomes effectively spin-independent.\ For example of $^{87}$Rb in the hyperfine state of $F$ $=$ $1$, $a_0$ $=$ $101.8$ and $a_2$ $=$ $100.4$ $a_B$ \cite{Stamperkurn2013} where $a_B$ is Bohr radius.\ The wavefunction in general can be expressed as
\bea
|\Psi\rangle=\sum_{s_1,s_2,...s_N}\psi_{s_1,s_2,...s_N}(\vec{x})|s_1,s_2,...,s_N\rangle,
\eea
where $\vec{x}$ $=$ $(x_1,x_2,...,x_N)$ denotes the spatial distributions along with the spin configurations of $N$ bosons, that is $|s_1,s_2,...s_N\rangle$ $\equiv$ $|\vec{s}\rangle$.\ 

The total wavefunction must be symmetric under interchange of any two particles.\ Therefore it is sufficient to first consider the region of $x_1$ $<$ $x_2$ $<$ $...$ $<$ $x_N$, and all others can be obtained via symmetry.\ For the spin part, we consider some degenerate and normalized spin configuration state $|\chi\rangle$.\ Later we will show that it does not matter which $|\chi\rangle$ we start with.\ Since TG gas limit suggests of null wavefunctions when $x_i$ approaches $x_j$, we can construct the symmetrized spatial part of the wavefunction $|\Psi'\rangle$ $=$ $\psi_{\vec{n}}^{sym}(\vec{x})|\chi\rangle$ in terms of the eigenfunctions $\phi_{n_j}(x_j)$ of the noninteracting fermions in a harmonic trap as
\bea
\psi_{\vec{n}}^{sym}(\vec{x})&=&\frac{1}{\sqrt{N!}}\mathbb{A}[\phi_{n_1}(x_1),\phi_{n_2}(x_2),...,\phi_{n_N}(x_N)]\nonumber\\
&\times&{\rm sgn}(x_2-x_1)\times{\rm sgn}(x_3-x_2) ...\nonumber\\
&\times&{\rm sgn}(x_N-x_{N-1}),\label{wf}
\eea
where $\mathbb{A}$ is anti-symmetrizer (equivalent to a Slater determinant), sgn is the sign function to satisfy the bosonic symmetry, and the factor of $\sqrt{N!}$ guarantees the normalization of the wavefunction.\ The eigenfunctions $\phi_n(x)$ (also dimensionless form of $\phi_n(y)$) in a harmonic trap are
\bea
\phi_n(y)&=&\frac{1}{\sqrt{2^n n!}}\frac{1}{\pi^{1/4}}H_n(y)e^{-y^2/2},~y\equiv x/x_{ho},
\eea
where $H_n$ are Hermite polynomials with the trap frequency $\omega$, the atom mass $m$, and $x_{ho}$ $\equiv$ $\sqrt{\hbar/(m\omega)}$.\

The above can be considered as permutations of the orbitals $\vec{n}$ $=$ ($n_1$, $n_2$, $...n_N$), and note that it should not be mistaken for permutations of $\vec{x}$ for here we have chosen the ordered region $x_1$ $<$ $x_2$ $<$ $...$ $<$ $x_N$.\ To access the regions other than $x_1$ $<$ $x_2$ $<$ $...$ $<$ $x_N$ of the wavefunction, we can construct them via permutations of the orderings of $\vec{x}$.\ There should be a total of $N!$ regions related to the symmetric group S$_N$.\ For $N$ $=$ $3$ as an example, we have $|\Psi'\rangle$ for the original ordered region, and a permutation of first two particles $P_{12}|\Psi'\rangle$ $=$ $\psi_{\vec{n}}^{sym}(\vec{x})P_{12}|\chi\rangle$ accesses the spatial region $x_2$ $<$ $x_1$ $<$ $x_3$ giving the new spin configuration $P_{12}|\chi\rangle$ for this region.\ Other four regions can be constructed by permutations of $P_{23}$, $P_{12}P_{23}$, $P_{23}P_{12}$, and $P_{12}P_{23}P_{12}$ on $|\chi\rangle$.\ Similar treatments of constructing the wavefunctions have been used in investigating quantum magnetism \cite{Deuretzbacher2008, Deuretzbacher2014} and magnetic correlations \cite{Volosniev2014} in strongly interacting 1D spinor gases.

Note that the ground state energy for such $N$ bosons occupying the lowest $N$ orbitals $\vec{n}$ $=$ ($0$, $1$, $...N-1$) is $E$ $=$ $N^2\hbar\omega/2$.\ Below we proceed to calculate the density matrix using the wavefunction we formulate here including the spin degree of freedom.

%%%%%%%%%%%%%%%%%%%%%%%%%%%%%%%%%%%%%%%%%%%%%%%%%%%%%%%%%%%%%%%%%%%%%%%%%%%%%%%%
\section{Density matrix}
Here we provide the formalism to calculate the single-particle density matrix of a spin-1 Bose gas, which reads
\bea
\rho(x,x')=N\sum_{\vec{s}}\int d\bar{x}\psi_{\vec{s}}^*(x,\bar{x})\psi_{\vec{s}}(x',\bar{x}),
\eea
where $\bar{x}$ $=$ $(x_2,x_3,...,x_N)$, and a factor of $N$ shows up for the other $N$ possible choices of $x$ and $x'$.\ We consider only the region of $x$ $<$ $x'$ which should be symmetric to $x$ $>$ $x'$, and also the ordering of $x_2$ $<$ $x_3$ $...$ $<$ $x_N$.\ For $N$ $=$ $3$ as an example we have the spin function overlaps in $\rho(x,x')$, for a given choice of $|\chi\rangle$,
\bea
x<x_2<x_3,&&~ x'<x_2<x_3,~\langle\chi|E|\chi\rangle=1,\nonumber\\
         ,&&~ x_2<x'<x_3,~\langle\chi|P_{12}|\chi\rangle,\nonumber\\
         ,&&~ x_2<x_3<x',~\langle\chi|P_{23}P_{12}|\chi\rangle,\nonumber\\
x_2<x<x_3,&&~ x_2<x'<x_3,~\langle\chi|E|\chi\rangle=1,\nonumber\\
         ,&&~ x_2<x_3<x',~\langle\chi|P_{12}^{-1}P_{23}P_{12}|\chi\rangle,\nonumber\\
x_2<x_3<x,&&~ x_2<x_3<x',~\langle\chi|E|\chi\rangle=1,
\eea
where the first two columns represent the integral regions of $x$ and $x'$, and the third column is the spin function overlap in this region.\ $E$ is the identical permutation operator, and we note that $P_{12}^{-1}P_{23}P_{12}$ $=$ $P_{13}$ and $P_{23}P_{12}$ $=$ $P_{123}$.\ After averaging over a set of spin states $|\chi\rangle$ (to be specified below), we then have the single-particle density matrix,
\bea
\rho(x<x')&=& 3\times 2\times\left\{\int_{x<x'<x_2<x_3}1\right.\nonumber\\
&+&\int_{x<x_2<x'<x_3}\frac{{\rm Tr}_\chi(P_{12})}{{\rm Tr}_\chi(E)}\nonumber\\
&+&\int_{x<x_2<x_3<x'}\frac{{\rm Tr}_\chi(P_{123})}{{\rm Tr}_\chi(E)}+\int_{x_2<x<x'<x_3}1\nonumber\\
&+&\left.\int_{x_2<x<x_3<x'}\frac{{\rm Tr}_\chi(P_{13})}{{\rm Tr}_\chi(E)}+\int_{x_2<x_3<x<x'}1\right\}\nonumber\\
&\times&\psi_{\vec{n}}^{sym*}(x,x_2,x_3)\psi_{\vec{n}}^{sym}(x',x_2,x_3)dx_2dx_3,
\eea
where the multiplication factor of $2$ in the above is due to the contribution of the integral region $x_2$ $>$ $x_3$.\ This region can be assessed by interchanging $x_2$ and $x_3$ using $P_{23}$, where the spin function overlaps are equivalent to those with $x_2$ $<$ $x_3$.\ The trace (Tr) of the permutation operator over $|\chi\rangle$ is just the summation of the spin function overlaps over all spin configurations, that is Tr$_{\chi}(P_{ij...k})$ $=$ $\sum_{\chi}$ $\langle\chi |P_{ij...k}|\chi\rangle$.\ To account for all the spin configurations, we have made an average over all the spin function overlaps in each integral with the total number of states given by Tr$_{\chi}(E)$.

To proceed and from now on we shall limit ourselves to the specific sector of total $S_z$ $\equiv$ $\sum_{i=1}^N \langle \hat{S}_z^i\rangle$ $=$ $0$, where $\hat{S}_z^i$ is $z$-component spin operator for $i$th particle.\ Then we have two possible sets of spin configurations, $\{|\chi_1\rangle^P\}$ $=$ $|000\rangle$, and $\{|\chi_2\rangle^P\}$ $=$ $|+-0\rangle$, $|+0-\rangle$, $|-+0\rangle$, $|-0+\rangle$, $|0+-\rangle$, $|0-+\rangle$, composing of six permuted states, which make up of seven states in total.\ We denote ($-$, $0$, $+$) as three quantum numbers ($-1$, $0$, $1$) for a spin-1 boson, and the superscript $P$ for a permuted set of spin configurations.\ Then the density matrix becomes
\bea
\rho(x<x')&=& 3\times 2\times\left\{\int_{x<x'<x_2<x_3}1+\int_{x<x_2<x'<x_3}\frac{1}{7}\right.\nonumber\\
&+&\int_{x<x_2<x_3<x'}\frac{1}{7}+\int_{x_2<x<x'<x_3}1\nonumber\\
&+&\left.\int_{x_2<x<x_3<x'}\frac{1}{7}+\int_{x_2<x_3<x<x'}1\right\}\nonumber\\
&\times&\psi_{\vec{n}}^{sym*}(x,x_2,x_3)\psi_{\vec{n}}^{sym}(x',x_2,x_3)dx_2dx_3,
\eea
where we note that Tr$_\chi(P_{12})$ $=$ Tr$_\chi(P_{13})$.\ The permutations of $P_{12}$, $P_{13}$, and $P_{23}$ are conjugate with each other to form a conjugacy class, therefore they have the same trace.\ In the $S_3$ permutation group, $E$, $P_{12}$, and $P_{123}$ form three classes.\ Their traces can be explicitly done and they are also listed in the Appendix A.

In general for $x$ $<$ $x'$, we have
\begin{widetext}
\bea
\rho(x<x')&=&N!\bigg\{\int_{x<x'<x_2...<x_N}1+\int_{x<x_2<x'...<x_N}\frac{w_{2N}}{w_N}+\int_{x<x_2<x_3<x'...<x_N}\frac{w_{3N}}{w_N}+...\nonumber\\
&+&\int_{x_2<x<x'...<x_N}1+\int_{x_2<x<x_3<x'...<x_N}\frac{w_{2N}}{w_N}+...+\int_{x_2<x_3...<x_N<x<x'}1\bigg\} \psi_{\vec{n}}^{sym*}(x,\bar{x})\psi_{\vec{n}}^{sym}(x',\bar{x})d\bar{x},\label{main_eq}
\eea
\end{widetext}
where the averaged spin function overlaps are given by $\frac{w_{jN}}{w_N}$, with
\bea
w_{jN}\equiv{\rm Tr}_{\chi}(P_{12...j}),\label{spin_weight}
\eea
the trace of $P_{12...j}$, averaged by
\bea
w_{N}\equiv{\rm Tr}_{\chi}(E),
\eea
the total number of states given $|\chi\rangle$.\ The spin function overlaps can be calculated using the conjugacy class $G$ of symmetric group $S_N$ where we demonstrate up to $N$ $=$ $6$ in the Appendix A.\ In addition the identity relation of $\sum_G{\rm Tr}_{\chi}G$ $=$ $N!$, the order of $S_N$, is useful to check on the calculation of various spin function overlaps.\ Note that the integral regions of permuted $\bar{x}$ are identical, therefore the density matrix $\rho(x<x')$ has $N(N+1)/2$ distinguished ones.

Here we show how we calculate the general spin function overlaps in the sector of $S_z$ $=$ $0$.\ The spin configurations $\{|\chi_i\rangle^P\}$ in general involves $|00...0\rangle$ and $n$ pairs of $(+-)$, that is $|++--00...0\rangle$ with $n$ $=$ $2$ for example.\ For $\{|\chi_i\rangle^P\}$ $=$ $|00...0\rangle$, the trace of any permutation operator is $1$.\ For the spin configuration of $n$ pairs of $(+-)$, the trace of $P_{12...j}$ has three contributions.\ One contribution is that the first $j$ entries are $(0)$'s, where the spin configuration can be shown as 
\bea
|\underbrace{00...0}_{j}\underbrace{00...0}_{N-2n-j}\underbrace{+...+}_{n}\underbrace{-...-}_{n}\rangle.\nonumber
\eea
The trace is $(N-j)!/[(n!)^2(N-2n-j)!]$ since this is the number of states obtained by permuting the rest of ($N-2n-j$) $(0)$'s and $n$ $(+)$'s and $(-)$'s.\ The other two contributions are for $j$ $(+)$'s or $(-)$'s, which is $(N-j)!/[(n-j)!n!(N-2n)!]$, the number of states obtained by permuting the rest of ($n-j$) $(+)$'s or $(-)$'s, $n$ $(-)$'s or $(+)$'s, and $(N-2n)$ $(0)$'s.\ In general we have the traces for the conjugacy classes of $P_{12...j}$ as
\bea
w_{jN}&\equiv&\sum_{n=0}^{\frac{N}{2}~{\rm or}~\frac{N-1}{2}}\left[\frac{(N-j)!}{(n!)^2(N-2n-j)!}\right.\nonumber\\
&+&\left.\frac{2(N-j)!}{(n-j)!n!(N-2n)!}\right],
\eea
for even or odd $N$ with the upper limit of summation as $N/2$ or $(N-1)/2$ respectively.\ We note that all the arguments of the factorials should be equal and larger than zero.\ The total number of states is then calculated as 
\bea
w_N\equiv\sum_{n=0}^{\frac{N}{2}~{\rm or}~\frac{N-1}{2}} \frac{N!}{(n!)^2(N-2n)!}.
\eea
In the same conjugacy class of $S_N$, the spin function overlaps are identical if the permutations are in the same class.\ The values of $w_{jN}$ can be found in Appendix A.

The momentum distributions of 1D Bose gas in TG limit is then numerically integrated based on Eq. (\ref{main_eq}) which we will demonstrate in Sec. V.\ The definition of the momentum distribution is
\bea
\rho(p)=\frac{1}{2\pi}\int_{-\infty}^\infty dx \int_{-\infty}^\infty dx' e^{ip(x-x')}\rho(x,x'),
\eea
where we let $\hbar$ $=$ $1$.\ Next we are interested in deriving the asymptotic forms in large momentum limit, which show $1/p^4$ decay and can be measured in experiments.
%%%%%%%%%%%%%%%%%%%%%%%%%%%%%%%%%%%%%%%%%%%%%%%%%%%%%%%%%%%%%%%%%%%%%%%%%%%%%%%%
\section{Large \texorpdfstring{$\boldsymbol{p}$}{p} expansion in \texorpdfstring{$\boldsymbol{\rho(p)}$}{rhop}}
Here we show the analytical result of a large $p$ asymptotic in the one-body momentum distribution of 1D TG Bose gas.\ It is well known that this large asymptotic shall show the universal $1/p^4$ dependence for a Bose gas with two-body contact interactions \cite{Minguzzi2002, Olshanii2003, Xu2015}.\ This universal asymptotic has recently drawn a lot of attentions in deriving the energetics (so called Tan's relation) \cite{Tan2008} and universal properties of the two-component Fermi gas \cite{Braaten2008-1, Braaten2008-2, Werner2009, Zhang2009}, and in extending Tan's relations in one dimension \cite{Barth2011}.\ Before we present the general expression of the large $p$ asymptotic for arbitrary $N$, we show the results of $N$ $=$ $2$ and $3$ for demonstration. 

%%%%%%%%%%%%%%%%%%%%%%%%%%%%%%%%%%%%%%%%%%%%%%%%%%%%%%%%%%%%%%%%%%%%%%%%%%%%%%%%
\subsection{\texorpdfstring{$\boldsymbol{N=2}$}{N}}
We first recall of $\rho(x<x')$ for two bosons in Eq. (\ref{main_eq}) that
\bea
\rho(x<x')&=&2!\left[\int_{x<x'<x_2}1+\int_{x<x_2<x'}\frac{w_{22}}{w_2}\right.\nonumber\\
&+&\left.\int_{x_2<x<x'}1\right]dx_2\psi_{\vec{n}}^{sym*}(x,x_2)\psi_{\vec{n}}^{sym}(x',x_2).\nonumber\\
\eea
If we replace the coefficient from the averaged spin function overlap ($w_{22}/w_2$ $=$ $1/3$) by $(-1)$, we end up with $\rho(x,x')$ equivalent to the results of the noninteracting fermions, therefore $\rho(p)$ $\propto$ $e^{-(px_{ho})^2}$.\ Subtracting this vanishing part of $e^{-(px_{ho})^2}$ in large $p$ limit, we have
\bea
\rho(p)&\underset{p\rightarrow\infty}{=}&\frac{2!}{2\pi}\int_{-\infty}^\infty dxdx'e^{ip(x-x')}\int_{x<x_2<x'}dx_2\big(\frac{w_{22}}{w_2}+1\big)\nonumber\\
&\times&\psi_{\vec{n}}^{sym*}(x,x_2)\psi_{\vec{n}}^{sym}(x',x_2)+(x'<x),\nonumber\\
&\underset{p\rightarrow\infty}{=}&\frac{1}{2\pi}\int_{-\infty}^\infty dxdx'e^{ip(x-x')}\bigg[-\frac{4}{3}\int_{x<x_2<x'}dx_2\bigg]\nonumber\\&\times&
\left| \begin{array}{cc}
\phi_0(x) & \phi_1(x) \\
\phi_0(x_2) & \phi_1(x_2) \end{array} \right|
\left| \begin{array}{cc}
\phi_0(x') & \phi_1(x') \\
\phi_0(x_2) & \phi_1(x_2) \end{array} \right|\nonumber\\
&+&(x'<x),
\eea
where we have used Eq. (\ref{wf}), and the second term represents the part of integrals for $x'$ $<$ $x$.\ 

Let $\bar{y}$ $=$ $x$ $-$ $x_2$ and $\bar{y}'$ $=$ $x'$ $-$ $x_2$ with $d\bar{y}$ $=$ $dx$ and $d\bar{y}'$ $=$ $dx'$, we have
\bea
\rho(p)&\underset{p\rightarrow\infty}{=}&\frac{1}{2\pi}\frac{-4}{3}\left[\int_{-\infty}^0 d\bar{y}\int_{0}^\infty d\bar{y}'+\int_{-\infty}^0 d\bar{y}'\int_{0}^\infty d\bar{y}\right]\nonumber\\
&\times& e^{ip(\bar{y}-\bar{y}')}\int_{-\infty}^\infty dx_2
\left| \begin{array}{cc}
\phi_0(\bar{y}+x_2) & \phi_1(\bar{y}+x_2) \\
\phi_0(x_2) & \phi_1(x_2) \end{array} \right|\nonumber\\
&\times&
\left| \begin{array}{cc}
\phi_0(\bar{y}'+x_2) & \phi_1(\bar{y}'+x_2) \\
\phi_0(x_2) & \phi_1(x_2) \end{array} \right|.
\eea
Since $p$ is large, $\bar{y}$ and $\bar{y}'$ are necessarily small so we can expand the determinants by Taylor expansion.\ Also since the zeroth order expansion gives null results, we keep the nonvanishing first-order expansion, which are the order of $\bar{y}$ and $\bar{y}'$ in the integrals.\ We then have
\bea
\rho(p)&\underset{p\rightarrow\infty}{=}&\frac{1}{2\pi}\frac{-4}{3}\left[\int_{-\infty}^0 d\bar{y}\int_{0}^\infty d\bar{y}'+\int_{-\infty}^0 d\bar{y}'\int_{0}^\infty d\bar{y}\right]\nonumber\\
&\times& \bar{y}\bar{y}' e^{ip(\bar{y}-\bar{y}')}
\int_{-\infty}^\infty dx_2
\left| \begin{array}{cc}
\phi_0'(x_2) & \phi_1'(x_2) \\
\phi_0(x_2) & \phi_1(x_2) \end{array} \right|^2,
\eea 
where the prime on the eigenfunctions means the derivative.\ 

We then proceed to solve the above integrals by the integration by parts.\ For brevity we just demonstrate the integrals of $\bar{y}$ and $\bar{y}'$,
\bea
\int_{-\infty}^0 d\bar{y}\int_{0}^\infty d\bar{y}' e^{ip(\bar{y}-\bar{y}')}\bar{y}\bar{y}'=\frac{-1}{p^4},\label{p4}
\eea
where we have imposed the conditions of negligible contribution at the infinite boundary.\ Note that the part of $(\bar{y}\leftrightarrow \bar{y}')$ gives the same result and put the above back to $\rho(p)$, we have
\bea
\rho(p)&\underset{p\rightarrow\infty}{=}&\frac{4/3}{2\pi}\frac{2}{p^4}\int_{-\infty}^\infty dx_2\left| \begin{array}{cc}
\phi_0'(x_2) & \phi_1'(x_2) \\
\phi_0(x_2) & \phi_1(x_2) \end{array} \right|^2,\nonumber\\
&=&\frac{4/3}{2\pi}\frac{2}{p^4}\int_{-\infty}^\infty \frac{e^{-2y^2}dy}{\pi(x_{ho})^3}\left| \begin{array}{cc}
-y & \sqrt{2}(1-y^2) \\
1 & \sqrt{2} y \end{array} \right|^2,
\eea
where we substitute $x_2$ by $y$ $=$ $x_2/x_{ho}$ in $\phi_n$.\ Further we use the dimensionless $\hbar\bar{k}$ $=$ $\bar{p}$ replacing of $p/\sqrt{m\hbar\omega}$, and we have
\bea
\frac{\rho(\bar{k})}{x_{ho}}&\underset{\bar{k}\rightarrow\infty}{=}&\frac{4/3}{2\pi}\frac{2}{\bar{k}^4}\int_{-\infty}^\infty\frac{2e^{-2y^2}dy}{\pi},\nonumber\\
&=&\frac{4/3}{2\pi}\frac{2}{\bar{k}^4}\frac{2}{\pi}\sqrt{\frac{\pi}{2}}=\frac{0.34}{\bar{k}^4}.
\eea
%%%%%%%%%%%%%%%%%%%%%%%%%%%%%%%%%%%%%%%%%%%%%%%%%%%%%%%%%%%%%%%%%%%%%%%%%%%%%%%%
\subsection{\texorpdfstring{$\boldsymbol{N=3}$}{N}}
Toward the general expression of large $p$ asymptotic for arbitrary $N$, we show one more example of three bosons.\ Recall again of Eq. (\ref{main_eq}), we have six integral regions in $\rho(x<x')$,
\bea
&&3!\bigg[\int_{x<x'<x_2<x_3}1+\int_{x<x_2<x'<x_3}\frac{w_{23}}{w_3}+\int_{x<x_2<x_3<x'}\frac{w_{33}}{w_3}\nonumber\\
&&+\int_{x_2<x<x'<x_3}1+\int_{x_2<x<x_3<x'}\frac{w_{23}}{w_3}+\int_{x_2<x_3<x<x'}1\bigg]\nonumber\\
&&\times dx_2dx_3\psi_{\vec{n}}^{sym*}(x,x_2,x_3)\psi_{\vec{n}}^{sym}(x',x_2,x_3),\label{eqN3}
\eea
where $w_{23}/w_3$ $=$ $w_{33}/w_3$ $=$ $1/7$.\ We can again subtract from the above the corresponding expression of the Fermi gas, which has only $e^{-(p x_{ho})^2}$ contribution in the large $p$ limit.\ Further, noting again that $\rho(p)$ has significant contributions only from the regions of $x$ $<$ $x_j$ $<$ $x'$ and $x'$ $<$ $x_j$ $<$ $x$ for all $x_j$ $\in$ $\bar{x}$, we then have the remaining terms only from $w_{23}/w_3$ $\to$ $w_{23}/w_3+1$ in Eq. (\ref{eqN3}) for $p$ $\to$ $\infty$,
\bea
\rho(p)&\underset{p\rightarrow\infty}{=}&\frac{1}{2\pi}\int_{-\infty}^\infty dxdx'e^{ip(x-x')}\bigg[-\frac{8}{7}\int_{x<x_2<x'<x_3}\nonumber\\
&-&\frac{8}{7}\int_{x_2<x<x_3<x'}\bigg]dx_2dx_3\nonumber\\
&\times&\left| \begin{array}{ccc}
\phi_0(x) & \phi_1(x) & \phi_2(x) \\
\phi_0(x_2) & \phi_1(x_2) & \phi_2(x_2) \\
\phi_0(x_3) & \phi_1(x_3) & \phi_2(x_3) \end{array} \right|\nonumber\\
&\times&\left| \begin{array}{ccc}
\phi_0(x') & \phi_1(x') & \phi_2(x') \\
\phi_0(x_2) & \phi_1(x_2) & \phi_2(x_2) \\
\phi_0(x_3) & \phi_1(x_3) & \phi_2(x_3) \end{array} \right|+(x'<x).
\eea

Again we change variables by $\bar{y}$ $=$ $x$ $-$ $x_2$, $\bar{y}'$ $=$ $x'$ $-$ $x_2$, or $x_2$ $\rightarrow$ $x_3$.\ Similar to the derivation of $N$ $=$ $2$, we expand the above to the first order of $\bar{y}$ and $\bar{y}'$, apply Eq. (\ref{p4}), and we have
\bea
\rho(p)&\underset{p\rightarrow\infty}{=}&\frac{1}{2\pi}\frac{8}{7}\frac{2}{p^4}\int_{-\infty}^\infty dx_2dx_3\Big|_{x_2<x_3}\nonumber\\&\times&\left[
\left| \begin{array}{ccc}
\phi_0'(x_2) & \phi_1'(x_2) & \phi_2'(x_2) \\
\phi_0(x_2) & \phi_1(x_2) & \phi_2(x_2) \\
\phi_0(x_3) & \phi_1(x_3) & \phi_2(x_3) \end{array} \right|^2\right.\nonumber\\
&+&\left.
\left| \begin{array}{ccc}
\phi_0'(x_3) & \phi_1'(x_3) & \phi_2'(x_3) \\
\phi_0(x_2) & \phi_1(x_2) & \phi_2(x_2) \\
\phi_0(x_3) & \phi_1(x_3) & \phi_2(x_3) \end{array} \right|^2\right].
\eea
The two terms inside the bracket are equivalent via $x_2\leftrightarrow x_3$, therefore we can combine these two terms, but now we can integrate over all $x_2$ and $x_3$.\ We then expand the determinant into minors, and integrate out either $x_2$ or $x_3$, obtaining
\bea
\rho(p)&\underset{p\rightarrow\infty}{=}&\frac{1}{2\pi}\frac{8}{7}\frac{2}{p^4}\int_{-\infty}^\infty dx\bigg[
\left| \begin{array}{cc}
\phi_1'(x) & \phi_2'(x) \\
\phi_1(x) & \phi_2(x) \end{array} \right|^2\nonumber\\&+&
\left| \begin{array}{cc}
\phi_0'(x) & \phi_2'(x) \\
\phi_0(x) & \phi_2(x) \end{array} \right|^2+
\left| \begin{array}{cc}
\phi_0'(x) & \phi_1'(x) \\
\phi_0(x) & \phi_1(x) \end{array} \right|^2\bigg].
\eea
From the above, it suggests that the integrals involve the determinants with a combination of every two eigenfunctions.\ To proceed, we put in the Hermite polynomials, and use the dimensionless $\bar{k}$,
\bea
\frac{\rho(\bar{k})}{x_{ho}}&\underset{\bar{k}\rightarrow\infty}{=}&\frac{2\cdot 8/7}{2\pi\bar{k}^4}\frac{1}{\pi}\int_{-\infty}^\infty dye^{-2y^2}\bigg[2+
\left| \begin{array}{cc}
-y & \frac{5y-2y^3}{\sqrt{2}} \\
1 & \frac{2y^2-1}{\sqrt{2}} \end{array} \right|^2\nonumber\\&+&
\left| \begin{array}{cc}
\sqrt{2}(1-y^2) & \frac{5y-2y^3}{\sqrt{2}} \\
\sqrt{2} y & \frac{2y^2-1}{\sqrt{2}} \end{array} \right|^2\bigg],\nonumber\\
&=&\frac{2\cdot 8/7}{2\pi\bar{k}^4}\frac{27\sqrt{\pi/2}}{4\pi}=\frac{0.98}{\bar{k}^4}.
\eea

%%%%%%%%%%%%%%%%%%%%%%%%%%%%%%%%%%%%%%%%%%%%%%%%%%%%%%%%%%%%%%%%%%%%%%%%%%%%%%%%
\subsection{General large \texorpdfstring{$\boldsymbol{p}$}{p} asymptotic}
In general for arbitrary $N$, we have
\bea
\frac{\rho(p)}{N}&&\underset{p\rightarrow\infty}{=}\frac{(N-1)!}{N!}\frac{2[1+w_{2N}/w_N]}{2\pi p^4}\sum_{j=2,3,...N}\nonumber\\
&&\int_{x_2<x_3...<x_N}d\bar{x}\left| \begin{array}{cccc}
\phi_{n_1}'(x_j) & \phi_{n_2}'(x_j) & ...\\
\phi_{n_1}(x_j) & \phi_{n_2}(x_j) &... \\
\vdots &... & \vdots \\
\phi_{n_1}(x_N) &... & \phi_{n_N}(x_N) \end{array} \right|^2,\nonumber\\\label{generalp}
\eea
where the factor of $(N-1)!$ is from the permutation of the ordering $x_2$ $<$ $x_3$ $...$ $<$ $x_N$.\ The above derivation of Eq. (\ref{generalp}) simply generalizes the previous case of $N$ $=$ $3$.\ We have expanded the determinants by Taylor expansion for $x$ $\approx$ $x_j$ $\approx$ $x'$, and used Eq. (\ref{p4}) for the Fourier transform.\ We have also rearranged the rows of the above determinant.\ Absorbing the factor of $(N-1)!$ to form all space integrals in $\bar{x}$, we have  
\bea
\rho(p)&\underset{p\rightarrow\infty}{=}&\frac{1}{(N-1)!}\frac{2[1+w_{2N}/w_N]}{2\pi p^4}\sum_{j=2,3,...N}\int_{-\infty}^\infty d\bar{x}\nonumber\\
&\times&
\left| \begin{array}{cccc}
\phi_{n_1}'(x_j) & \phi_{n_2}'(x_j) & ...\\
\phi_{n_1}(x_j) & \phi_{n_2}(x_j) &... \\
\vdots &... & \vdots \\
\phi_{n_1}(x_N) &... & \phi_{n_N}(x_N) \end{array} \right|^2.
\eea

From the orthogonality of the eigenfunctions $\phi_n$, finally we can reduce the above as
\bea
\rho(p)\underset{p\rightarrow\infty}{=}\frac{2[1+\frac{w_{2N}}{w_N}]}{2\pi p^4}\sum_{(n_i,n_j)}\int_{-\infty}^\infty dx
\left| \begin{array}{cc}
\phi_{n_i}'(x) & \phi_{n_j}'(x) \\
\phi_{n_i}(x) & \phi_{n_j}(x) \end{array} \right|^2,\nonumber\\\label{large_p}
\eea 
where $(n_i,n_j)$ denotes any possible pairs of $N$ eigenfunctions, and the factor $(N-1)!$ cancels out by $(N-1)$ summations of $x_j$ within each there are $(N-2)!$ copies of the minors.\ In the below we show specifically the results for finite number of atoms up to $N$ $=$ $6$ .
%%%%%%%%%%%%%%%%%%%%%%%%%%%%%%%%%%%%%%%%%%%%%%%%%%%%%%%%%%%%%%%%%%%%%%%%%%%%%%%%

For $N$ $=$ $4$, we have $10$ terms of integral regions.\ From Eq. (\ref{large_p}), we have
\bea
\frac{\rho(\bar{k})}{x_{ho}}\underset{\bar{k}\rightarrow\infty}{=}\frac{2\cdot 24/19}{2\pi\bar{k}^4}\frac{18.604}{\pi}=\frac{2.38}{\bar{k}^4}.
\eea
For $N$ $=$ $5$ and $6$, we have respectively
\bea
\frac{\rho(\bar{k})}{x_{ho}}\underset{\bar{k}\rightarrow\infty}{=}\frac{2\cdot 64/51}{2\pi\bar{k}^4}\frac{33.57}{\pi}=\frac{4.27}{\bar{k}^4},\\
\frac{\rho(\bar{k})}{x_{ho}}\underset{\bar{k}\rightarrow\infty}{=}\frac{2\cdot 180/141}{2\pi\bar{k}^4}\frac{53.93}{\pi}=\frac{6.98}{\bar{k}^4}.
\eea

%%%%%%%%%%%%%%%%%%%%%%%%%%%%%%%%%%%%%%%%%%%%%%%%%%%%%%%%%%%%%%%%%%%%%%%%%%%%%%%%
\section{Numerical results and discussions}
The momentum distribution of a 1D spinor Bose gas has been studied in some selected spin configurations \cite{Deuretzbacher2008}, which shows Friedel-like oscillations similar to the noninteracting fermions or broadened momentum distributions depending on the symmetry of the spin functions.\ Here we consider the spin-incoherent regime using Eq. (\ref{main_eq}), and demonstrate the momentum distribution of a 1D spin-1 Bose gas in TG limit.\ We use Monte Carlo (MC) integration method to derive $\rho(x,x')$, which is implemented in Linux system with message processing interface.\ To have a sense of the computation time, for $N$ $=$ $5$ with MC simulations of $M$ $=$ $1$e$7$ sets of random numbers, it requires about two days with $250$ parallel CPU cores.

%%%%%%%%%%%%%%%%%%%%%%%%%%%%%%%%%%%%%%%%%%%%%%%%%%%%%%%%%%%%%%
\begin{figure}[t]
\centering
\includegraphics[width=8.5cm,height=4.5cm]{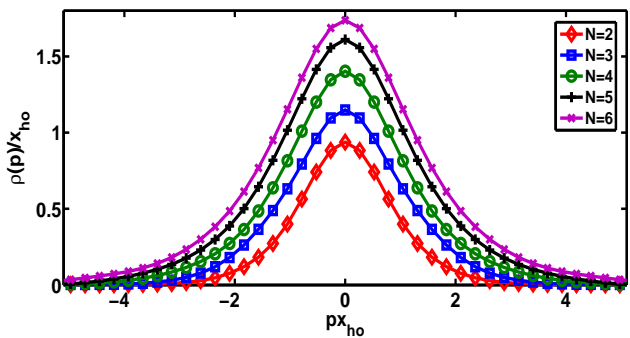}
\caption{(Color online) Momentum distributions of a spin-incoherent 1D spin-1 TG gas up to $N$ $=$ $6$ in the sector of $S_z$ $=$ $0$.\ Uniformly broadened distributions as $N$ grows are demonstrated, which are due to the spin function overlaps.}\label{fig1}
\end{figure}
%%%%%%%%%%%%%%%%%%%%%%%%%%%%%%%%%%%%%%%%%%%%%%%%%%%%%%%%%%%%%

%%%%%%%%%%%%%%%%%%%%%%%%%%%%%%%%%%%%%%%%%%%%%%%%%%%%%%%%%%%%%
\begin{figure}[b]
\centering
\includegraphics[width=8.5cm,height=4.5cm]{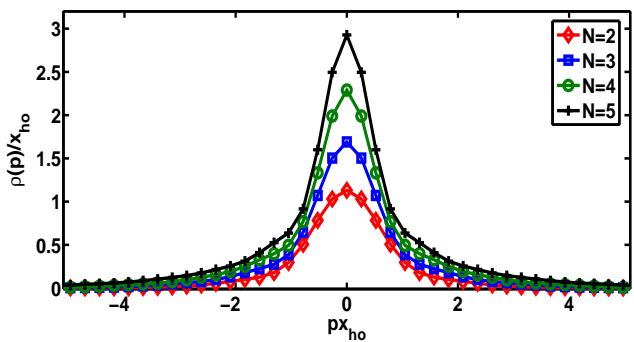}
\caption{(Color online) Momentum distributions of a 1D spinless TG gas up to $N$ $=$ $5$.\ The momentum distributions become sharper as $N$ increases.}\label{fig2}
\end{figure}
%%%%%%%%%%%%%%%%%%%%%%%%%%%%%%%%%%%%%%%%%%%%%%%%%%%%%%%%%%%%% 

In Fig. \ref{fig1}, the momentum distribution is demonstrated up to $N$ $=$ $6$.\ As the number of particles increases, its momentum distribution broadens almost uniformly in the peaks and the widths,\ This hugely contrasts with noninteracting fermions or spinless bosons, which have Friedel oscillations or narrower peaks at $p$ $=$ $0$ respectively.\ The uniformly broadened and structureless distribution is due to the spin function overlaps in the spin-incoherent regime that averages out the oscillatory features.\ The MC simulations in Fig. \ref{fig1} are $M$ $=$ $1$e$4$, $1$e$5$, $1$e$6$, $1.8$e$8$, $3.5$e$8$ for $N$ $=$ $2$ $-$ $6$.\ As a comparison, we show the results of spinless bosons (or equivalently fully spin polarized) in Fig. \ref{fig2} where we have a sharp momentum distribution as $N$ increases while the widths of the distributions stay almost the same.\ Therefore the spin-incoherent Bose gas has a broadened momentum distribution relative to the spinless case.\ At finite temperature, the momentum distribution would also be broadened.\ However, we shall see below that these two situations can be easily distinguished.

%%%%%%%%%%%%%%%%%%%%%%%%%%%%%%%%%%%%%%%%%%%%%%%%%%%%%%%%%%%%%%%%%%%%%%%%%%%%%%%%%%%%%%%%%%%%%%%%%%%%%%%%%%%%%%%%%%%%%%%%%%
\begin{figure}[t]
\centering
\includegraphics[width=8.5cm,height=4.5cm]{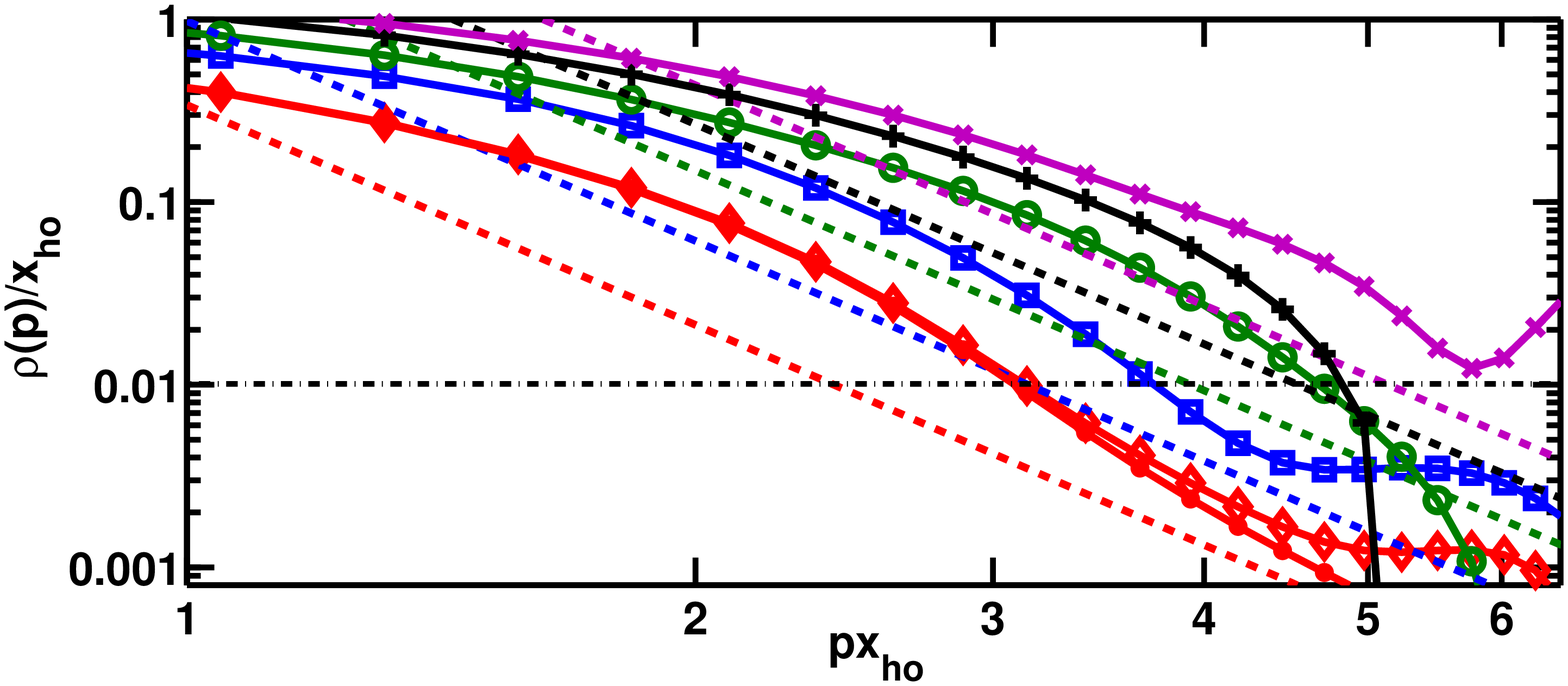}
\caption{(Color online) The asymptotics of large momentum distributions in Fig. (\ref{fig1}).\ Large momentum asymptotics are plotted in logarithmic scales and compared with analytic calculations (dash) in a 1D spin-1 TG gas.\ (a) The analytically derived asymptotics are $0.34/p^4$, $0.98/p^4$, $2.38/p^4$, $4.27/p^4$, and $6.98/p^4$ respectively for $N$ $=$ $2$ $-$ $6$.\ The result of $N$ $=$ $2$ (dash-$\bullet$) obtained by conventional numerical integration almost overlaps with the one by Monte Carlo integration, which asymptotically approaches the analytic curve in large $p$ limit as expected.\ The line symbols and colors follow Fig. \ref{fig1}, and a horizontal line of $10^{-2}$ is used to guide the eye for approximately the accuracy for the Monte Carlo results.}\label{fig3}
\end{figure}
%%%%%%%%%%%%%%%%%%%%%%%%%%%%%%%%%%%%%%%%%%%%%%%%%%%%%%%%%%%%%

%%%%%%%%%%%%%%%%%%%%%%%%%%%%%%%%%%%%%%%%%%%%%%%%%%%%%%%%%%%%%
\begin{figure}[b]
\centering
\includegraphics[width=8.5cm,height=4.5cm]{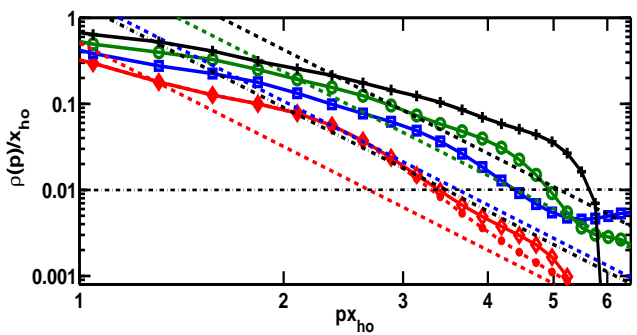}
\caption{(Color online) The asymptotics of large momentum distributions in Fig. (\ref{fig2}).\ Large momentum asymptotics are plotted in logarithmic scales and compared with analytic calculations (dash) in a 1D spinless TG gas.\ The analytically derived asymptotics are $0.51/p^4$, $1.715/p^4$, $3.77/p^4$, and $6.8/p^4$ respectively for $N$ $=$ $2$ $-$ $5$, and we also compare with Olshanii's calculation \cite{Olshanii2003} of large momentum result $1.45/p^4$ (dash-dot) for $N$ $=$ $5$.\ The result of $N$ $=$ $2$ (dash-$\bullet$) obtained by conventional numerical integration again overlaps with the one by Monte Carlo integration, which asymptotically approaches the analytic curve in large $p$ limit as expected.\ Similarly the line symbols and colors follow Fig. \ref{fig2}, and a horizontal line of $10^{-2}$ is also plotted for approximately the accuracy for the Monte Carlo results.}\label{fig4}
\end{figure}
%%%%%%%%%%%%%%%%%%%%%%%%%%%%%%%%%%%%%%%%%%%%%%%%%%%%%%%%%%%%%

In Figs. \ref{fig3} and \ref{fig4}, we show the large momentum asymptotics in logarithmic scales and compare with our analytic calculations in Sec. IV.\ The asymptotics for the spinless case can be obtained from Eq. (\ref{large_p}) by replacing $(1 + w_{2N}/w_{N})$ by $2$.\ The $N=2$ case can also be evaluated exactly.\ Our numerical results match well with the exact ones.\ However, we see that although the momentum distribution is approaching their asymptotic values, the differences between them remain rather large for $p x_{ho} \approx 3$.\ For larger $N$, one may expect that the asymptotics would take even larger $p$ to reach \cite{Olshanii2003}.\ Our numerical results become inaccurate for large $p$ for larger $N$'s.\ However, we can still make use of the asymptotics to improve our evaluation of the total kinetic energy of the system, as discussed below.\ For the spinless case, we also show the results from the asymptotic formula $\approx$ $0.1297 \times N^{3/2}/\bar{k}^4$ proposed by Olshanii and Dunjko \cite{Olshanii2003}.\ We see that in Fig. \ref{fig4}, for $N$ $=$ $5$, this value is much below the asymptotic that we obtain from (the modified) Eq. (\ref{large_p}).

\begin{figure}[t]
\centering
\includegraphics[width=8.5cm,height=4.5cm]{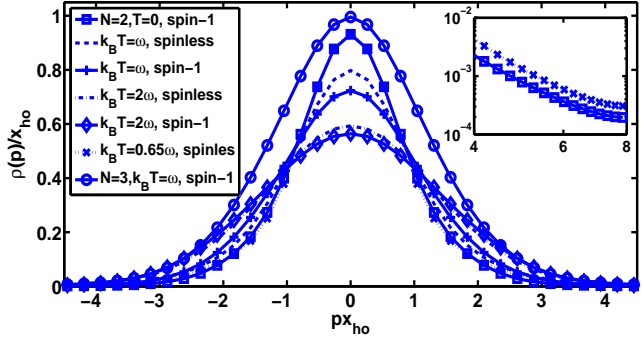}
\caption{(Color online) Momentum distributions comparisons of 1D spin-1 and spinless TG gas at finite temperature.\ The widths of the distributions broaden as the temperature increases for both the spin-1 and spinless case.\ To differentiate the momentum distributions of both, a fitting case of spinless bosons at $T$ $=$ $0.65\hbar\omega/k_B$ ($\times$) overlaps with spin-1 case at small $p$ while they differ in the large momentum limit (inset).}\label{fig5}
\end{figure}

At finite temperature, we show in Fig. \ref{fig5} that the momentum distributions are broadened as the temperature increases.\ The spinless case always has higher peaks than the spin-1 bosons at the same temperature.\ The case of $N$ $=$ $3$ is numerically averaged over all orbitals by $e^{-\beta E_s}$ $[\beta$ $\equiv$ $1/(k_BT)]$ at the cutoff of $E_s$ $=$ $6\hbar\omega$ where it shows no significant changes when including more orbitals.\ We also compare the spin-1 bosons at zero temperature with the spinless case at $T$ $=$ $0.65\hbar\omega/k_B$ where they almost overlap at the peaks while differ in the tails of the momentum distributions as shown in the inset.\  The spin-incoherent regime in spin-1 bosons can be distinguished from the spinless ones from the measurements of the momentum distributions and the system energies as well. 

%%%%%%%%%%%%%%%%%%%%%%%%%%%%%%%%%%%%%%%%%%%%%%%%%%%%%%%%%%%%%
%%%%%%%%%%%%%%%%%%%%%%%%%%%%%%%%%%%%%%%%%%%%%%%%%%%%%%%%%%%%%

The total energy ($\langle E_s \rangle$) of the system is the sum of the potential ($\langle V \rangle$) and kinetic ($\langle K \rangle$) contributions.\ For our 1D Bosonic system in the TG limit, the density distribution is identical with a Fermi gas.\ At zero temperature, we then have $\langle V \rangle$ $=$ $N^2 \hbar \omega / 4$.\ The kinetic energy can be obtained easily by considering the action of the Hamiltonian on the wavefunction at a point where all $x_j$'s are unequal.\ We see easily that $\langle K \rangle$ $=$ $N^2 \hbar \omega / 4$.\ The above is in accordance with the Virial theorem \cite{Werner2006, Werner2008}, $\langle K \rangle$ $=$ $\langle V \rangle$ $=$ $\langle E_s \rangle /2$.\ Accordingly, both the momentum distributions in Figs. \ref{fig1} and \ref{fig2} have the same value of $\int dp p^2 \rho(p)$.\ The sharper momentum distribution for the spinless case implies that $\rho(p)$ must be larger at larger $p$ than the spin incoherent case, as can be seen in Fig. \ref{fig3}.\ Correspondingly, while the momentum distribution of the spinless case broadens with increasing temperature, this broadened distribution is distinct from the broadening due to spin averaging as the total kinetic energy must be higher at finite temperature. 

We have also evaluated numerically the potential and kinetic energies by Monte Carlo integration.\ The potential energy is numerically derived by evaluating $\int dx x^2 \rho(x)$, which is always below the relative error $1.5\%$ to the exact value of $\langle V\rangle$.\ For $\langle K\rangle$ $\propto$ $\int dp p^2 \rho(p)$, it is more demanding of the accuracy in numerical calculations.\ We then attach the tails of our momentum distributions by the analytically derived asymptotics starting at around $px_{ho}$ $\approx$ $4.5$, which improves the relative error of $\langle K\rangle$ significantly to below $9\%$ for example in the case of spin-1 bosons.\ In Monte Carlo integration of our 1D spin-1 and spinless bosons, the integral boundaries is set to $x$ $=$ $\pm 4x_{ho}$ with $41\times 41$ meshes in $\rho(x,x')$.\ The results are convergent within $M$ $=$ $1$e$4$, $1$e$5$, $1$e$6$, $5$e$7$, $3.5$e$8$ for $N$ $=$ $2$ $-$ $6$, and note that the symmetry of $\rho(x,x')$ $=$ $\rho(x',x)$ can be used to double the $M$, thus reducing the computation time required for convergence.

%%%%%%%%%%%%%%%%%%%%%%%%%%%%%%%%%%%%%%%%%%%%%%%%%%%%%%%%%%%%%
\section{Conclusion}
In conclusion, we have investigated the properties of the spin-incoherent Luttinger liquid in a spin-1 Bose gas.\ The density matrix of such universal class can be calculated by the spin function overlaps from the highly degenerate spin configurations.\ We show that the spin function overlap is directly related to the traces of the conjugacy classes in the permutation groups.\ We also analytically derive the universal dependence of $1/p^4$ in large $p$ limit, and compare with the momentum distributions of a spin-1 Bose gas in TG limit using Monte Carlo integrations up to six bosons.\ The spin-incoherent Bose gas has a broadened momentum distribution which we can confirm and distinguish from the spinless case by measuring its momentum distribution and the total kinetic energy.\ The ultracold spinor Bose gas thus sets up a promising paradigm to realize this universal while different class of Luttinger liquid.

%%%%%%%%%%%%%%%%%%%%%%%%%%%%%%%%%%%%%%%%%%%%%%%%%%%%%%%%%%%%%
\section*{Acknowledgements}
This work is supported by the Ministry of Science and Technology, Taiwan, under Grant No. MOST-101-2112-M-001-021-MY3 and 104-2112-M-001-006-MY3.
%%%%%%%%%%%%%%%%%%%%%%%%%%%%%%%%%%%%%%%%%%%%%%%%%%%%%%%%%%%%%%%%%%%%%%%%%%%%%%%%%%%%%%%%%%%%%%%%%%%%%%%%%%%%%%%%%%%%%%%%%%
\appendix
\section{Traces of the conjugacy classes of \texorpdfstring{$\boldsymbol{S_N}$}{SN}}
Here we list the traces of the conjugacy classes relevant to the spin function overlaps.\ We note here that the number of conjugacy classes of $S_N$ is $2$, $3$, $5$, $7$, $11$ for $N$ $=$ $2-6$.

For $N$ $=$ $2$, we have $\{|\chi_1\rangle^P\}$ $=$ $|00\rangle$ and $\{|\chi_2\rangle^P\}$ $=$ $|+-\rangle$, $|-+\rangle$ respectively.\ In table I we have the traces in these bases for $E$ and $P_{12}$ corresponding to two conjugacy classes, and $G$ represents the group operators in relevant conjugacy classes in the calculations of spin function overlaps.

\begin{table}[ht]
\begin{tabular}[t]{c|cc}
\backslashbox{Class}{Traces} & Tr$_{\chi_1}(G)$ & Tr$_{\chi_2}(G)$\\
\hline
$E$                    & 1 & 2 \\
$P_{12}$             & 1 & 0 \\
\end{tabular}
\caption{Traces of the conjugacy classes of $S_2$.}
\end{table}

For $N$ $=$ $3$, we have $\{|\chi_{1,2}\rangle^P\}$ $=$ $|000\rangle$ and $\{|+-0\rangle^P\}$ respectively where $\{|+-0\rangle^P\}$ represent six different states by permutations.\ In table II we give the traces for the three conjugacy classes, $E$, $P_{12}$, and $P_{123}$, in these sets of states.

\begin{table}[ht]
\begin{tabular}[t]{c|cc}
\backslashbox{Class}{Traces} & Tr$_{\chi_1}(G)$ & Tr$_{\chi_2}(G)$\\
\hline
$E$                    & 1 & 6 \\
$P_{12}$               & 1 & 0 \\
$P_{123}$              & 1 & 0 \\
\end{tabular}
\caption{Traces of the conjugacy classes of $S_3$.}
\end{table}

For $N$ $=$ $4$, we have $\{|\chi_{1,2,3}\rangle^P\}$ $=$ $|0000\rangle$, $\{|+-00\rangle^P\}$, and $\{|+-+-\rangle^P\}$ respectively.\ In table III we give the traces for the four conjugacy classes, $E$, $P_{12}$, $P_{123}$, and $P_{1234}$, in these sets.\ The other conjugacy class involves $P_{12}P_{34}$ which we do not need in calculating the spin function overlap of Eq. (\ref{spin_weight}) due to the intended order of particle positions we choose in the first place.\ However it helps confirm our calculated traces from the identity relation, which we will demonstrate in the end of this appendix.

\begin{table}[ht]
\begin{tabular}[t]{c|ccc}
\backslashbox{Class}{Traces} & Tr$_{\chi_1}(G)$ & Tr$_{\chi_2}(G)$ & Tr$_{\chi_3}(G)$ \\
\hline
$E$                    & 1 & 12 & 6 \\
$P_{12}$               & 1 & 2  & 2 \\
$P_{123}$              & 1 & 0  & 0 \\
$P_{1234}$             & 1 & 0  & 0 \\
\end{tabular}
\caption{Traces of the conjugacy classes of $S_4$.}
\end{table}
%%%%%%%%%%%%%%%%%%%%%%%%%%%%%%%%%%%%%%%%%%%%%%%%%%%%%%%%%%%%%%%%%%%%%%%%%%%%%%%%

For $N$ $=$ $5$, we have $\{|\chi_{1,2,3}\rangle^P\}$ $=$ $|00000\rangle$, $\{|+-000\rangle^P\}$, and $\{|+-+-0\rangle^P\}$ respectively.\ In table IV we give the traces for the five conjugacy classes, $E$, $P_{12}$, $P_{123}$, $P_{1234}$, and $P_{12345}$, in these sets.

\begin{table}[htb]
\begin{tabular}[t]{c|ccc}
\backslashbox{Class}{Traces} & Tr$_{\chi_1}(G)$ & Tr$_{\chi_2}(G)$ & Tr$_{\chi_3}(G)$ \\
\hline
$E$                    & 1 & 20 & 30 \\
$P_{12}$             & 1 & 6 & 6   \\
$P_{123}$            & 1 & 2 & 0   \\
$P_{1234}$           & 1 & 0 & 0   \\
$P_{12345}$          & 1 & 0 & 0   \\
\end{tabular}
\caption{Traces of some conjugacy classes of $S_5$.}
\end{table}

For $N$ $=$ $6$, we have $\{|\chi_{1,2,3,4}\rangle^P\}$ $=$ $|000000\rangle$, $\{|+-0000\rangle^P\}$, $\{|+-+-00\rangle^P\}$, and $\{|+-+-+-\rangle^P\}$ respectively.\ In table V we give the traces for the six conjugacy classes, $E$, $P_{12}$, $P_{123}$, $P_{1234}$, $P_{12345}$, and $P_{123456}$, in these sets.

\begin{table}[ht]
\begin{tabular}[t]{c|cccc}
\backslashbox{Class}{Traces} & Tr$_{\chi_1}(G)$ & Tr$_{\chi_2}(G)$ & Tr$_{\chi_3}(G)$ & Tr$_{\chi_4}(G)$ \\
\hline
$E$                    & 1 & 30 & 90 & 20 \\
$P_{12}$             & 1 & 12 & 18 & 8  \\
$P_{123}$            & 1 & 6  & 0  & 2  \\
$P_{1234}$           & 1 & 2  & 0  & 0  \\
$P_{12345}$          & 1 & 0  & 0  & 0  \\
$P_{123456}$         & 1 & 0  & 0  & 0  \\
\end{tabular}
\caption{Traces of some conjugacy classes of $S_6$.}
\end{table}

Finally we demonstrate the details for the construction of conjugacy classes and the calculation of the traces in the class.\ Take four bosons in the symmetric group $S_4$ for example, the classification of the conjugacy classes of $S_N$ can be derived by a cycle decomposition that counts the number of unordered integer partitions.\ We then decompose four bosons as 
\bea
&&4,~3+1,~2+2,~2+1+1,~1+1+1+1,\nonumber\\
&&\rightarrow ~P_{1234},~P_{123},~P_{12}P_{34},~P_{12},~E,
\eea
which accounts for five conjugacy classes.\ The size of each conjugacy class can be calculated as $N!/(\Pi_j(j)^{a_j}a_j!)$ where we have $a_j$'s integer of $j$ in the unordered integer partitions.\ This calculation of the size can be also seen as $N$ permutations with $a_j$ times of $j$'s cycling and $a_j$ permutations of the groups $P_{12...j}$, $P_{j+1...2j}$, etc.\ Take $\{|\chi_3\rangle^P\}$ of $N$ $=$ $4$ as an example where we have three classes, $E$, $P_{12}$, and $P_{12}P_{34}$, that have nonvanishing traces.\ Their traces in this basis are $6$, $2$, and $2$ respectively with the sizes $1$, $6$, and $3$.\ Therefore the identity relation reads
\bea
\sum_G{\rm Tr}_{\chi}(G)=6\times 1+2\times 6+2\times 3=4!,
\eea
where the traces are verified.\ 

For the final demonstration of the identity relations in the symmetric group $S_5$, the classification of the conjugacy classes becomes
\bea
&&5,~4+1,~3+2,~3+1+1,~2+2+1,~2+1+1+1,\nonumber\\
&&1+1+1+1+1,\nonumber\\
&&\rightarrow P_{12345},~P_{1234},~P_{123}P_{45},~P_{123},~P_{12}P_{34},~P_{12},~E,
\eea
which accounts for seven conjugacy classes.\ We can verify the traces of each conjugacy classes by the identity relation.\ Take $\{|\chi_2\rangle^P\}$ of $N$ $=$ $5$ as an example, 
\bea
\sum_G{\rm Tr}_{\chi}(G) = 20\times 1+6\times 10 + 2\times 20 = 5!,
\eea
where the sizes of $P_{12}$ and $P_{123}$ are $10$ and $20$ respectively.\ Lastly for $\{|\chi_3\rangle^P\}$, we have
\bea
\sum_G{\rm Tr}_{\chi}(G) = 30\times 1+6\times 10 + 2\times 15 = 5!,
\eea
where the size of $15$ belongs to $P_{12}P_{34}$ and its trace is $2$.\ The above examples show the identity relation between the total traces of the conjugacy classes and the order of $S_N$.
%%%%%%%%%%%%%%%%%%%%%%%%%%%%%%%%%%%%%%%%%%%%%%%%%%%%%%%%%%%%%%%%%%%%%%%%%%%%%%%%%%%%%%%%%%%%%%%%%%%%%%%%%%%%%%%%%%%%

%%%%%%%%%%%%%%%%%%%%%%%%%%%%%%%%%%%%%%%%%%%%%%%%%%%%%%%%%%%%%%%%%%%%%%%%%%%%%%%%%%%%%%%%%%%%%%%%%%%%%%%%%%%%%%%%%%%%
\end{document}